\documentclass[floatfix,amsmath,amssymb,superscriptaddress,prl,preprint,nobibnotes,nofootinbib]{revtex4}

\usepackage{graphicx}
\usepackage{amsmath}
\usepackage{amssymb}
\usepackage{epsfig}
\usepackage{wasysym}
\usepackage{bbm}
\usepackage[colorlinks,citecolor=blue]{hyperref}
\usepackage{tabularx}

\renewcommand{\v}[1]{{\boldsymbol{#1}}}
\renewcommand{\c}[1]{{\cal #1}}

\newcommand{\imth}{\hspace{1pt}\mathrm{i}\hspace{1pt}}
\newcommand{\dif}{\mathrm{d}}
\newcommand{\fourfoldangle}{{90$^\circ$}}
\newcommand{\action}{S}
\newcommand{\spinfive}{\phi}

\newcommand{\Ref}[1]{Ref.~\onlinecite{#1}}
\newcommand{\Eq}[1]{equation~(\ref{#1})}
\newcommand{\Fig}[1]{Fig.~\ref{#1}}

\newcommand{\ie}{{\emph{i.e.~}}}
\newcommand{\eg}{{\emph{e.g.~}}}

\newcommand{\etal}{{\emph{et al.~}}}

\newcommand{\<}{\langle}
\renewcommand{\>}{\rangle}

\newcommand{\Tr}{{\rm Tr}}

\newcommand{\p}{\partial}

\newcommand{\ra}{\rightarrow}
\newcommand{\s}{{\sigma}}

\begin{document}

\title{
% Is FeSe a nematic quantum paramagnet?
Nematicity and quantum paramagnetism in FeSe
}

%% Notice placement of commas and superscripts and use of &
%% in the author list
%% Notice placement of commas and superscripts and use of &
%% in the author list

\author{Fa Wang}
\affiliation{International Center for Quantum Materials, School of Physics, Peking University, Beijing 100871, China.}
\affiliation{Collaborative Innovation Center of Quantum Matter, Beijing, China.}

\author{Steven A. Kivelson}
\affiliation{Department of Physics, Stanford University, Stanford, California 94305, USA.}

\author{Dung-Hai Lee}
% \email{dunghai@berkeley.edu}
\affiliation{Department of Physics, University of California, Berkeley, CA 94720, USA.}
\affiliation{Materials Sciences Division, Lawrence Berkeley National Laboratory, Berkeley, CA 94720, USA.}

\begin{abstract}
%\noindent\textbf{
In common with other iron-based high temperature superconductors, 
FeSe exhibits a transition to a ``nematic'' phase below 90Kelvin 
in which the 
crystal rotation symmetry is spontaneously broken.  
However, the absence of strong low-frequency magnetic fluctuations near 
or above 
the transition 
has been interpreted as implying the primacy of orbital ordering. 
In contrast, we establish that quantum fluctuations of spin-1 local moments 
with strongly frustrated exchange interactions can lead to a nematic quantum paramagnetic phase 
consistent with the observations in FeSe. 
We show that this phase is a fundamental expression of the existence of a Berry's phase 
associated with the topological defects of a N\'eel antiferromagnet, 
in a manner analogous to that which gives rise to valence bond crystal order for spin 1/2 systems. 
We present an exactly solvable model realizing the nematic quantum paramagnetic phase, 
discuss its relation with the spin-1 $J_1-J_2$ model, 
and construct a field theory of the Landau-forbidden transition between the N\'eel state and 
this nematic quantum paramagnet.
\end{abstract}
%}

\maketitle

Insulating quantum paramagnets (PM) are magnetic systems
with only short-range antiferromagnetic (AF) correlations even at zero temperature.
Examples of quantum paramagnetic states include valence bond solids (VBS),
symmetry protected topological states (SPTs), and 
spin liquids.
Interest in quantum paramagnets was greatly intensified 
following the proposal\cite{Anderson-Science87} (since shown to be incorrect) 
that the parent insulator of the cuprate high temperature superconductors (HTSC)
might be a spin liquid.
Because of the proposed relevance to the cuprates,
special attention has been paid to
square lattice $S=1/2$ systems,
and in particular to the $J_1-J_2$ Heisenberg model,
\begin{equation}
H=J_1\sum_{\<ij\>}\v S_i\cdot\v S_j +J_2\sum_{\<\<jk\>\>}\v S_j\cdot\v S_k.
\label{j1j2}
\end{equation}
where $\<ij\>$ and $\<\<jk\>\>$ denote nearest-neighbor (NN) and second neighbor pairs of sites.
Numerical studies have shown\cite{Kivelson-etal-PRB90,JiangHC-Balents-PRB12,Gong-PRL14}
that the ground state of \Eq{j1j2}
has N\'eel AF order for $0\le J_2/J_1\lesssim 0.4$ and stripe AF order [see
Fig.~2(a,b)]
% \Fig{fig2}(a,b)]
for $0.6\lesssim J_2/J_1$.
Both these phases have gapless $S=1$ (spin wave) excitations.
For $0.4\lesssim J_2/J_1\lesssim 0.6$ the ground state
appears to have no magnetic order.
However, there remain important unresolved issues 
concerning the precise character of this phase or phases;
at least for $0.5 \lesssim J_2/J_1\lesssim 0.6$ there is fairly compelling evidence of
translation symmetry breaking VBS order
with an energy gap for spin-1 excitations.

The curious fact that upon restoring the spin rotation symmetry 
the breaking of spatial translation symmetry follows rests on a uniquely quantum mechanical effect 
associated with the Berry's phase
of the monopole events.\cite{Haldane-PRL88,Read-Sachdev-PRL89}
A monopole (anti-monopole) is a space-time event [see
Fig.~1(a)]
% \Fig{fig1}(a)]
across which the ``skyrmion number'' (see the caption of
Fig.~1)
% \Fig{fig1})
changes by $+1(-1)$.
The proliferation of monopoles randomizes the N\'eel order parameter
hence causes the system to become a PM.
In \Ref{Haldane-PRL88},
it was 
%shown 
argued 
that monopole events contribute
to the path integral with a phase factor
that depends on the monopole's spatial location [see
Fig.~1(b,c)].
% \Fig{fig1}(b,c)].
\begin{center}
\begin{figure}
\includegraphics[scale=0.8]{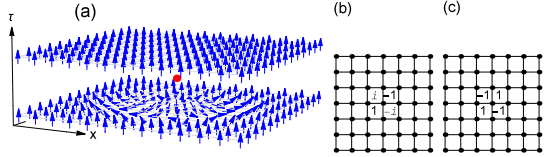}
\caption{{\bf The monopole phase factors}.
A monopole is a singular configuration of the N\'eel order parameter
whose direction is depicted by an unit vector $\hat{n}$
at each point in space and time.
The N\'eel order parameter configurations before and after
a monopole event differ topologically
-- the skyrmion number changes by one.
To understand the skyrmion number imagine assigning an unit vector to each point of
a two dimensional space subject to periodic boundary conditions.
Such an assignment is a ``map'' from a torus to the unit sphere $S^2$.
These maps can be grouped into topological classes, 
where only maps within the same class can be smoothly deformed into one another.
An integer, namely the number of times, and sense, 
the image of the torus wrap around the unit sphere, characterizes each class.
The skyrmion number is given by
$q_s={\frac{1}{4\pi}}\int \dif x \dif y~ \hat{n}\cdot\p_x\hat{n}\times\p_y\hat{n}$.
A close inspection of the arrow patterns in panel (a) reveal 
that before the monopole event (marked by the red dot),
the skyrmion number is $-1$, while that after the monopole event it is $0$.
An analogous figure can be drawn for an anti-monopole across which the skyrmion number jumps by $-1$.
Haldane showed that associated with each charge $q_m$ monopole event 
there is a phase factor, $\eta^{q_m}_{\cal R}$,
which enters the path integral
over all possible $\hat{n}$ configurations in space and (imaginary) time
\cite{Haldane-PRL88}.
This phase factor depends on the spatial location of the monopole
core,
which it is natural to associate with the center of a lattice plaquette, ${\cal R}$.
There is some arbitrariness in the choice of $\eta_{\cal R}$, 
but a consistent pattern for spin-1/2 on a square lattice 
is shown in panel (b) and for spin-1 in panel (c).
}
\label{fig1}
\end{figure}
\end{center}

Spin models based on local moments, such as the $J_1-J_2$ model,
have found renewed applications in
the young field of iron-based superconductivity.
The iron-based HTSCs have layers of Fe$^{2+}$ ions
which form a square lattice at high temperatures.
Many 
experimental and theoretical studies have 
concluded 
that, with the possible exception of heavily phosphorous doped members of the 122 family, 
the electronic correlations in the iron-based materials, in particular in the iron chalcogenides, 
are 
%quite 
strong\cite{Basov-NatPhys09,SiQimiao-NJP09,Kotliar-NatMat11}.
Moreover, relatively large local magnetic moments in many of these materials 
have been inferred\cite{KimYJ-PRB11}. For 
FeSe 
the itinerant electrons only give rise to tiny Fermi pockets\cite{Borisenko-PRB14,Takahashi-PRL14,ColdeaAI-PRB15,DingH-FeSe-15}. 
Thus, it is plausible that the magnetism of FeSe can be addressed using a spin model as a starting point.

For many iron-based materials, depending on the doping level,
there is 
 ``stripe'' AF order at low temperatures
[see
Fig.~2(a,b)].
% \Fig{fig2}(a,b)].
Due to the breaking of crystal rotation symmetry the stripe order is accompanied by
a tetragonal to orthorhombic lattice distortion. %Interestingly i
In some cases,
\eg in electron-doped Ba-122 materials,
the lattice distortion can exist without stripe order [see
Fig.~3(a)].
% \Fig{fig3}(a)].
In \Ref{ChuJH-Science12}
it was shown that 
the distortion is driven by electronic nematicity rather than a lattice (phonon) instability.
It has been argued
that such nematicity reflects an underlying stripe ordering tendency, 
and the reason it can exist without the magnetic order is
because thermal fluctuations of the continuously varying spin orientation are more severe
than of the discrete nematic director\cite{Chandra-Coleman-Larkin,FangC-PRB08,XuCK-PRB08}.
\begin{center}
\begin{figure}
\centering
\includegraphics[scale=0.8]{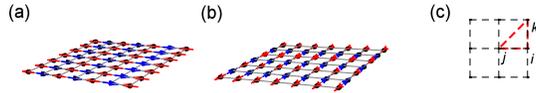}
\caption{
{\bf The stripe AF and the interactions in $H_K$:}
Panels (a) and (b) show two degenerate versions of a stripe AF ground state 
corresponding to the two possible directions of the ordering vector, 
where the arrows schematically represent the ordered magnetic moments on the Fe sites. 
Panel (c) illustrates the three-site interaction
corresponding to each individual projection operator from \Eq{proj}.
}
\label{fig2}
\end{figure}
\end{center}

A potential problem with this perspective is the thermal evolution observed in FeSe crystals,
in which nematicity onsets at $T_{\rm nem}\sim$90Kelvin, but
there is no magnetic ordering down to the lowest measured temperatures
suggesting the possibility of a zero temperature nematic quantum PM phase
[Fig.~3(b)].
% [\Fig{fig3}(b)].
This fact coupled with the absence of any observed enhancement of 
the low frequency magnetic fluctuations at $T_{\rm nem}$\cite{Buchner-NatMat15,Meingast-PRL15}
has led many to concluded that the nematicity in FeSe is driven by orbital ordering.

Certainly, it is clear that
within a local moment picture the lack of magnetic order
implies that the spin-spin interactions must be highly frustrated;  this is true regardless of 
what causes the nematic ordering.
Thus in the following we ask, 
``Can frustrated spin interactions alone drive a nematic quantum PM state?'' 

We consider models in which each Fe$^{2+}$ possesses a % local moment associated a 
localized 
spin 1
(\eg the $S=1$ version of the model in \Eq{j1j2}).
Given the strength of Hund's coupling and the crystal-field splittings, 
$S=1$ is a reasonable possibility for the spin of Fe$^{2+}$ ions.
However it is important to keep in mind that FeSe is not an insulator, 
and  such localized models neglect the effects of itinerant carriers.
We shall return to this point later.

\begin{center}
\begin{figure}
\centering
\includegraphics[scale=0.8]{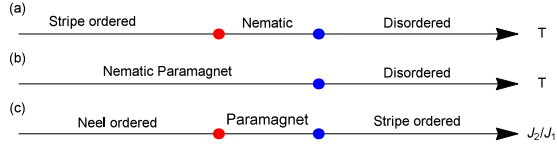}
\caption{
{\bf Schematic phase diagrams:}
(a) The thermal phase diagram  of many iron-based superconducting materials.
The blue and red dots represent two distinct phase transitions:
as a function of decreasing $T$,
 the discrete crystal rotation symmetry is spontaneously broken at the blue point and continuous spin rotation symmetry at the red.
(b) A likely thermal phase diagram %for the FeSe crystal.
of FeSe.
(c) The zero temperature phase diagram of the spin-1 $J_1$--$J_2$ model in \Eq{j1j2} as a function of $J_2/J_1$.
}
\label{fig3}
\end{figure}
\end{center}

First we demonstrate the existence of a nematic quantum PM phase
in an exactly solvable Hamiltonian.
Consider a square lattice of $S=1$ spins interacting via
the short-range, spin rotationally invariant Hamiltonian
\begin{equation}
H_K=K\sum_{\<ijk\>}%P_{3;jik}
P_3(\v S_i+\v S_j+ \v S_k),
\label{proj}
\end{equation}
where $K>0$ and $\sum_{\<ijk\>}$ sums over all
elementary triangles of sites [see
Fig.~2(c)],
% \Fig{fig2}(c)]
and
where $\overline{ji}$ and $\overline{ik}$ are NN bonds
and $\overline{jk}$ is a next-NN bond.
Here
$P_3(\v S) =  (1/720) S^2(S^2-2)(S^2-6)$ is the projection operator onto $S=3$.
We note that because it involves spin-1 operators,
\Eq{proj} possesses a global spin SO(3) [rather than SU(2)] symmetry.
In addition it possesses all crystalline symmetries of the square lattice
(\ie translation and point group symmetries).
Moreover,
there are two degenerate ground states which can be constructed exactly as follows:
Any closed loop ${\cal C}_{i_1,...,i_n}$ on the lattice can be thought of as a spin-1 chain.
The famous AKLT (Affleck-Kennedy-Lieb-Tasaki) 
state\cite{AKLT} 
of such chains %is proportional to
%the sum over all pairs of nearest-neighbor bonds, $\overline{ji}$,
%of projection operators, $P_2(\v S_i+\v S_j)$, onto spin 2;
%its unique ground state 
can be written %(among other representations)
as the matrix product state
% \begin{equation}
$
|{\cal C}_{i_1,...,i_n}\>=\sum_{m_{i_1}=-1}^1...\sum_{m_{i_n}=-1}^1
\Tr\left[A(m_{i_1})...A(m_{i_n})\right]|m_{i_1},...,m_{i_n}\>,
$
% \label{aklt}
% \end{equation}
where $m_{i_k}=-1,0,1$ is the $S_z$ quantum number of the spin on site $i_k$,
and the matrices $A(m)$ are
% \begin{equation}
$
A(\pm 1)=(i\s_y\mp 1)/2\sqrt{2},~
A(0)=\s_z/2,$ 
with $\s_{y,z}$ being the Pauli matrices.
% \label{A}
% \end{equation}
We identify the two adjacent sites $i_k$ and $i_{k+1}$ in an AKLT loop as
an ``AKLT-entangled pair'', and graphically represent it
by a blue bond connecting $i_k$ and $i_{k+1}$ in
Fig.~4.
% \Fig{fig4}.
The two ground-states of $H_K$ are constructed as
the direct product of AKLT loop states on all the loops
made by connecting nearest-neighbor sites in the $x$ direction, 
$|{\rm X}\>=\prod_{y=1}^N|{\cal C}_{(1,y),(2,y),...,(N,y)}\>$,
or in the $y$ direction, 
$|{\rm Y}\>=\prod_{x=1}^N|{\cal C}_{(x,1),(x,2),...,(x,N)}\>$,
where $(x,y)$ with $x,y=1,\dots,N$ labels the sites of a $N\times N$ square lattice.
The graphical representations of $|{\rm X}\>$ and $|{\rm Y}\>$ are 
shown 
in
Fig.~4.
% \Fig{fig4}.
Because the maximum total spin of an AKLT-entangled pair is $S = 1$\cite{AKLT},
it follows that the maximum spin component of a triplet of spins
$\<ijk\>$ containing at least one AKLT-entangled pair is $S = 2$.
From this it follows trivially that $H_K$ annihilates $|{\rm X}\>$ and $|{\rm Y}\>$;
moreover, since $H_K$ is positive semidefinite, this proves that they are ground states.
\begin{center}
\begin{figure}
\includegraphics[scale=0.8]{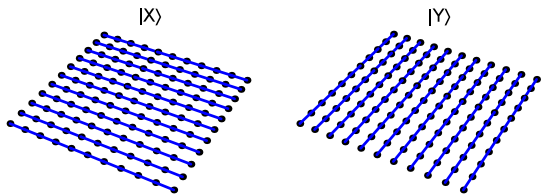}
\caption{
{\bf The nematic paramagnetic ground states of $H_K$ from \Eq{proj}.}
The solid blue lines connect AKLT-entangled pairs discussed in the main text.}
\label{fig4}
\end{figure}
\end{center}

It is only slightly more difficult to see that any other state constructed as
a direct product of AKLT-loop states has positive energy:
since each site can be AKLT entangled with at most two other sites,
and since there are four elementary triangles per site,
in any zero energy state exactly two of the sites must be AKLT-entangled,
and each entangled pair must form an edge of four distinct elementary triangles.
The two states $|{\rm X}\>$ and $|{\rm Y}\>$ are the only ones that satisfy this constraint.
For instance, if the blue bonds in
Fig.~4
% \Fig{fig4}
make a turn,
an elementary triangle with its vertex at the corner of the turn
will not contain any AKLT-entangled pair.
While we do not have a general proof that no more complex ground states exist,
explicit finite cluster diagonalization
(up to system size $4\times 4$) suggest that there are none.

From the known properties of AKLT states,\cite{AKLT,Arovas-Auerbach-Haldane}
it follows immediately that $|{\rm X}\>$ and $|\rm{Y}\>$ are gapped PM states
with exponentially falling spin-spin correlations
which break crystal rotation symmetry:
$G_X(\vec R_{ij})\equiv \<{\rm X}|\v S_i\cdot\v S_j |{\rm X}\> =
9\cos({\vec R_{ij}\cdot \vec Q_N})\delta(\vec R_{ij}\cdot \hat y) e^{-|\vec R_{ij}\cdot \hat x|/\xi_0}$
with $\vec Q_N = \pi(\hat x +\hat y)$ and $\xi_0 = 1/\ln(3)$ in units in which the lattice constant is 1.
The gap implies that the nematic PM phase is perturbatively stable.
However, the asymptotic form of $G$ is non-generic -- this reflects the fact
that $H_K$ lies on a ``disorder line\cite{Emery}''
where, although there is no associated thermodynamic non-analyticity,
the oscillatory character of the short-range order changes.

Turning to less ``reverse engineered'' models, we discuss the
very interesting numerical (density matrix renormalization group) study
by Jiang \etal\cite{JiangHC-PRB09} of the spin-1 version of
the $J_1-J_2$ model in \Eq{j1j2}.
They found that as a function of $J_2/J_1$ there is
an intermediate paramagnetic phase (characterized by gapped $S=1$ excitations)
between the N\'eel and stripe ordered phases [see
Fig.~3(c)].
% \Fig{fig3}(c)].
Moreover the N\'eel and stripe order parameters vanish %es 
continuously
as $J_2/J_1$ approaches the respective critical values
from the magnetically ordered side (see Fig.~4 of \Ref{JiangHC-PRB09}).
In the supplementary information we perform exact diagonalization on small lattices 
and demonstrate the adiabatic connectivity
of the ground-state phases of $H_K$ (\Eq{proj}) 
and the $J_1-J_2$ model when $J_2/J_1$ falls within the PM regime. 
In addition we compare the 
excitations in different spin and momentum sectors.

To obtain an analytic understanding of the nematic quantum PM phase,
we consider a field theory description valid in the neighborhood of
a continuous or weakly first order 
quantum phase transition
from the N\'eel phase to a nematic PM phase.
Because the unbroken symmetries of the N\'eel and nematic PM phases do not have a subgroup relationship,
classical Landau theory would imply that such a continuous phase transition is forbidden.
However as pointed out in \Ref{deconfined}, when the topological defects of one order
carry the quantum number of the other order a continuous transition becomes possible.

According to \Ref{Haldane-PRL88}, in the path integral describing
the quantum fluctuations of the $S=1$ N\'eel antiferromagnet in 2D,
the weight associated with each field configuration is determined by the usual non-linear sigma model (analogous to the first term in \Eq{nlsm}),
but there is also an additional Berry's phase factor.
For a ``charge'' $q_m$ monopole
(which causes the skyrmion number to jump by $q_m$)
centered on plaquette ${\vec {\cal R}}$ (which designates a point on the dual lattice)
this phase factor 
(up to a global ``gauge'' ambiguity) is 
$\eta_{\vec {\cal R} }= e^{i q_m(\vec Q_N\cdot \vec {\cal R}) }$.
[see
Fig.~1(c)].
% \Fig{fig1}(c)].
Rotation by
{\fourfoldangle}
about a lattice site transforms $\eta_{\vec {\cal R}}\to -\eta_{\vec {\cal R}}$
but keeps $q_m$ invariant.
So long as this symmetry is preserved, the Feynman amplitude of 
any field configuration with an odd $q_m$ cancels that associated with  
the configuration rotated by
{\fourfoldangle}.
As a result odd $q_m$ monopoles cannot proliferate,
although even $q_m$ monopoles can.
Consequently, states can be classified by a conserved skyrmion number parity
$(-1)^{q_s}$. 
It turns out that 
the two-fold degeneracy associated with opposite skyrmion parity 
accounts for the two nematic ground states depicted in 
Fig.4.
% \Fig{fig4}.}
If we introduce a perturbation that breaks the crystal
{\fourfoldangle}
rotation symmetry,
the absolute value of the Feynman amplitudes associated with monopoles sitting at the $+1$ and $-1$ locations in
Fig.~1(c)
% \Fig{fig1}(c)
do not need to be the same anymore, hence 
their Feynman amplitudes no longer cancel
which means the odd $q_m$ monopoles can proliferate rendering the PM state non-degenerate.
This observation justifies identifying the two-fold degeneracy in the symmetric system
with spontaneous breaking of
{\fourfoldangle}
rotation symmetry.

In the following discussions we will use Euclidean space-time
and denote the imaginary time as $\tau$.
Consider a 4-component real vector field of unit norm,
$\hat{\Omega}(x,y,\tau)=(\v\Omega,\Omega_4)$
with $|\hat \Omega|^2=1$,
whose first three components, $\v {\Omega}(x,y,\tau)$ are the N\'eel order parameter
and $\Omega_4$ is the Ising-like nematic order parameter.
For example, for a spin model on a square lattice,
we can take $\Omega_4(x_i,y_i,\tau)\propto \<\v S_i\cdot\v S_{i+\hat{x}}-\v S_i\cdot\v S_{i+\hat{y}}\>$.
The non-linear sigma model action we shall consider is
\begin{equation}
% \begin{aligned}
\action=
% &
\int \dif^2x \dif\tau~\left[{\frac{1}{2g}}
|{\partial_\mu \hat \Omega}|^2 + V(\Omega_4^2)\right]
% \\ &
+\imth {\frac{\Theta}{2\pi^2}}\int \dif^2x \dif\tau~\epsilon^{abcd}\Omega_a{\partial_x\Omega_b} {\partial_y\Omega_c} {\partial_\tau\Omega_d},
% \end{aligned}
\label{nlsm}
\end{equation}
where $\Theta=\pi$.
This field theory has been introduced in \Ref{Abanov-Wiegmann-NPB00} and \Ref{XuCK-Ludwig}
which discuss %es
 the effects of topological terms on the spectrum and phases of non-linear sigma models.
For our purposes a derivation of \Eq{nlsm} is given in the supplementary information.
Note that this particular topological term is possible only in spatial dimension $d=2$ with a $N=4$ component order parameter.
Here $V$ is an anisotropy term that favors the first three components of $\hat{\Omega}$,
(for example $V=\Delta[\Omega_4]^2$), hence \Eq{nlsm} has  $O(3)\times Z_2$ symmetry.
Consider a  monopole configuration centered on the origin, $(x,y,\tau)=(0,0,0)$.
Because of the anisotropy term,
$\hat{\Omega}$ on a two-sphere ($S^2$) in space-time far away from the center of the monopole
is largely constrained to lie in the space spanned by the first three components of $\hat{\Omega}$,
\ie $\hat{\Omega} \approx (\hat n,0)$.
Topologically the configurations of $\hat{n}$ on the two-sphere are classified by the skyrmion number
which, in this case, is equal to the monopole charge $q_m$.
However 
such far field configuration can be compatible with two different $\Omega_4$ orientations
in the monopole core. 
Consider the following configurations
which reduce to
the same $q_m=1$ configuration in $\hat{n}$
far from monopole the center
\begin{equation}
\hat{\Omega}^{(\pm)}(x,y,\rho)=\left(\sin\phi_\pm(\rho)~\hat{n}(x,y,\rho),\cos\phi_\pm(\rho)\right),
\label{twomono}
\end{equation}
where $\rho=\sqrt{x^2+y^2+\tau^2}/R$,
$\phi_+(\rho)={\frac{\pi}{2}}\tanh(\rho)$, and
$\phi_-(\rho)=\pi(1-{\frac{1}{2}}\tanh(\rho))$.
Here $R$ is the size of the ``monopole core''.
$\hat{\Omega}^{(\pm)}$ both describe $q_m=1$ monopoles
but with $\hat\Omega=(\v 0,\pm 1)$ in the monopole core.
It is straightforward to show that
$\exp\{i {\frac{\Theta}{2\pi^2}}\int d^2x d\tau~\epsilon^{abcd}\Omega_a{\partial_x\Omega_b} {\partial_y\Omega_c} {\partial_\tau\Omega_d}\}
=e^{\pm i\pi/2}=\pm i$
for these two types of monopole.
A similar discussion holds for the $q_m=-1$ monopole.
It is easy to generalize the above argument to $q_m=2$ monopoles and obtain the 
corresponding 
phase factors $e^{\pm i\pi}=-1$.
A consequence of the $Z_2$ (nematic) symmetry is that the Feynman amplitudes of the two monopoles described above have the same absolute values.

These monopoles represent events (as a function of imaginary time) 
at  which the skyrmion number changes.
We can thus think of the quantum disordered phase as an interacting fluid of skyrmions and anti-skyrmions.
Because of the destructive interference between %the
the two types of  $q_m=1$ monopoles discussed above,
events in which the skyrmion number changes by one are forbidden --
the net skyrmion number is thus conserved modulo 2 and
the Hilbert space breaks into an even and an odd sector.
Since the stiffness constant $1/g$ renormalizes to zero in the quantum disordered state
and since there is a non-zero (possibly small) density of skyrmions and anti-skyrmions,
the ground-state energy in these two sectors should be equal in the thermodynamic limit.
This is the ground-state degeneracy associated with the breaking of C$_4$ symmetry to C$_2$.
In the supplementary information we discuss the effects of an explicit C$_4$ $\ra$ C$_2$ symmetry breaking field. 

One can also arrive at the PM phase in
Fig.~3(c)
% \Fig{fig3}(c)
from the stripe ordered side by proliferating the monopole of the stripe order parameter.
Generalizing the calculation of \Ref{Haldane-PRL88} we find
the Berry's phase factor associated with the stripe monopole is trivial.
Consequently the charge $\pm 1$ monopole can proliferate rendering the resulting quantum disordered state non-degenerate.
Here, because the symmetries of the two phases have subgroup relationship, a continuous transition is allowed in the Landau theory.
We expect such transition to %have
be in 
the $O(3)$ universality class.

Our key theoretical conclusion is that for a spin-1 model on a square lattice,
any ``intermediate'' quantum disordered phase between the N\'eel and stripe ordered magnetic phases
is likely to be a nematic PM.
This is consistent with the numerical evidence for the $J_1-J_2$ model [\Eq{j1j2}]. 
Because the nematic quantum PM phase straddles closely between the Neel and nematic phases, we expect it to have low-lying spin excitations near both momenta $(\pi,0)$ and $(\pi,\pi)$.
The nematic order considered here might be thought of as ``vestigial'' order\cite{Nie-Kivelson} left behind
when quantum fluctuations have restored spin-rotational symmetry.

Before speculating on the relevance of these results to the interesting case of FeSe,
it is necessary to comment on the effects of itinerant carriers.
ARPES and STM experiments show very small electron and hole pockets for FeSe\cite{ColdeaAI-PRB15,DingH-FeSe-15,Matsuda-PNAS14}
and at the same time quantum oscillation associated with these tiny pockets are seen\cite{TerashimaT-PRB14,ColdeaAI-PRB15}. 
Because the latter experiment has very stringent requirement on the long life time of the quasiparticles
we take 
it 
as an indication that the coupling between the itinerant carriers and the local moments is weak. 
This phase with decoupled magnetic and itinerant carrier degrees of freedom  is analogous to the antiferromagnetic metal phase in heavy fermion systems.
However due to the itinerant carriers
the nematic phase can no longer be characterized as having a spin-gap, and close enough to criticality,
the universality class of the quantum critical point is likely to be altered. 
In addition, FeSe is at best ``quasi-2D,'' which is to say
that 
it is 
ultimately three dimensional, and this, too, 
will alter the nature of the phase transitions,
but not necessarily affect the sequence of ordered phases.

Since the iron-based superconductors involve multiple $3d$ orbitals it has been long suspected that orbital ordering is the cause of nematicity. Althoug we focus on the nematicity caused by frustrated magnetism, we do not intend to imply that  orbital degrees of freedom play no role at all. As is well known regardless of the driving mechanism the nematic order parameter induces orbital splitting and lattice distortion. Therefore cooperative effect involving many different degrees of freedom can well be necessary to describe the system behavior quantitatively.In addition the so far ignored spin orbit coupling is ultimately necessary for the nematicity in the magnetic degrees of freedom  to induce the band splitting observed by ARPES and anisotropic magnetic susceptibility observed by NMR..etc.

With these caveats, the present results suggest that
it may be possible to view the nematic phase in the 
FeSe as being driven primarily by frustrated magnetism. 
The fact that the underlying nematic quantum PM is gapped implies
that there need not be any enhancement of 
the low energy magnetic fluctuations associated with the proximity of stripe long range order of the sort 
that has been detected by NMR in other Fe-based materials.   
Recent neutron scattering experiments\cite{ZhaoJun-FeSe-15} show
relatively short-range magnetic correlations but with reasonably large intensity
(\ie with substantial magnitudes of the fluctuating moments)
at the stripe ordering wavevector at low energies. 
Consequently, the lack of any strong evidence of slow magnetic fluctuations in NMR\cite{Meingast-PRL15}
and the short-range of the magnetic correlations seen in neutron scattering appear to be
entirely consistent with a magnetic origin of nematicity.  
As discussed previously we expect the nematic
PM phase to also exhibit relatively low energy spin excitations at the Neel ordering wavevector.
The nematic quantum PM state consists of a stacking of spontaneously formed 
spin-1 AF chains. It can be viewed as a ``weak'' symmetry protected topological state.
A potentially testable consequence of this observation is 
that gapless spin 1/2 excitations can arise on certain surfaces and domain boundaries.
In addition we expect nonmagnetic impurities to cut the chain and induce low lying spin excitations in the PM gap. 
This can be detected by electron spin resonance in a manner analogous to that done for spin-1 chains\cite{PhysRevLett.67.1614}.
Further experimental studies of nematic order and fluctuations in these materials --
including elasto-resistance, nuclear quadrupole resonance, and Raman scattering studies --
as well more complete neutron scattering studies of the magnetic fluctuations would clearly be helpful.

Finally after the submission of our paper, 
related theoretical studies appeared\cite{Hirschfeld-FeSe-15,SiQimiao-FeSe-15}
including one based on an itinerant electron picture of the nematicity in FeSe\cite{Chubukov-FeSe-15}.

% \begin{addendum}
% \item[Acknowledgements:]
\textbf{Acknowledgements:}
We thank Hongchen Jiang and Tao Xiang for useful discussions.
FW was supported by National Key Basic Research Program of China (Grant No. 2014CB920902)
and National Science Foundation of China (Grant No. 11374018).
SAK was supported in part by the U.S. Department of Energy, Office of Science, Basic Energy Sciences, Materials Sciences and Engineering Division, grant DE-AC02-76SF00515 at Stanford.
DHL was supported by the U.S. Department of Energy, Office of Science, Basic Energy Sciences, Materials Sciences and Engineering Division, grant
DE-AC02-05CH11231.
DHL and SAK would like to thank KITP for hospitality, 
supported in part by the National Science Foundation under Grant No. NSF PHY11-25915, 
where the collaboration started.

\appendix

\section{The Spin-1 N\'eel-Nematic Transition as a double Spin-1/2 N\'eel-VBS Transition}
\label{sec:spin-half}

It has been proposed that
a Landau-forbidden continuous quantum phase transition
between N\'eel order and valence bond solid
can happen in spin-1/2 systems on square lattice.
This transition can be described by a
% $O(3)\times C_{4v}$
non-linear sigma model (NL$\sigma$M)
with a Wess-Zumino-Witten(WZW) term and certain anisotropy terms\cite{Tanaka-Hu-PRL05,Senthil-Fisher-PRB05}.
The action of this model reads
\begin{equation}
\action_{\frac{1}{2}}[\hat \spinfive]=
\action_{O(3)\times C_{4v}}[\hat \spinfive]
-2\pi \imth
\frac{3}{8\pi^2}\int \dif u  \dif^2 x \dif\tau\,
\epsilon^{a b c d f}
\spinfive_a
\partial_x \spinfive_b
\partial_y \spinfive_c
\partial_\tau \spinfive_d
\partial_u \spinfive_f.
\label{eq:O5-WZW}
\end{equation}

Here the 5-component ``superspin'' $\hat{\spinfive}\propto (n_x,n_y,n_z,v_x,v_y)$
consists of the N\'eel order parameters $\v n=(n_x,n_y,n_z)\sim (-1)^{x+y}\v S_{(x,y)}$,
and the two columnar VBS order parameters $\v v=(v_x,v_y)$ where
\begin{subequations}
\begin{eqnarray}
v_x & \sim &  (-1)^{x}(\v S_{(x,y)}\cdot \v S_{(x+1,y)}-\v S_{(x,y)}\cdot\v S_{(x-1,y)}),\\
v_y & \sim &  (-1)^{y}(\v S_{(x,y)}\cdot \v S_{(x,y+1)}-\v S_{(x,y)}\cdot\v S_{(x,y-1)}).
\end{eqnarray}
\end{subequations}
$u$ is the auxiliary dimension for defining the WZW term.
$\action_{O(3)\times C_{4v}}$
is the non-topological part of the non-linear sigma model action
with $O(3)\times C_{4v}$ symmetry.
This action includes the stiffness terms such as
$\int  \dif^2 x \dif\tau\,
\left (
\frac{1}{2g_{n}}|\partial_{\mu} \v n|^2 +
\frac{1}{2g_{v}}|\partial_{\mu} \v v|^2
\right )
$. In addition it contains anisotropy terms that favor the  N\'eel order parameters over the VBS order parameters. Moreover, among the VBS order parameters there are terms that favor the columnar VBS over the plaquette VBS order. Examples of such anisotropy terms include
$\Delta\cdot (\spinfive_4^2+\spinfive_5^2)$  (where $\Delta>0$)
and
$U\cdot (\spinfive_4^2-\spinfive_5^2)^2$ (where $U<0$).
%The purpose of the anisotropy terms is to ensure 
In particular, $\Delta>0$ insures that 
the low energy physics is described by the fluctuations of the N\'eel order parameters.

Spin-1 can be viewed as two spin-1/2s coupled by strong ferromagnetic(FM) interaction.
We thus consider two copies of the action \Eq{eq:O5-WZW} labeled by superscripts $^{(1)}$
and $^{(2)}$,
\begin{equation}
\action_{1}[\hat\spinfive^{(1)},\hat\spinfive^{(2)}]
=\action_{\frac{1}{2}}[\hat \spinfive^{(1)}]
+\action_{\frac{1}{2}}[\hat \spinfive^{(2)}]
+
\int \dif^2 x \dif\tau\,
\left (
J_n \v n^{(1)}\cdot \v n^{(2)}
+
J_v \v v^{(1)}\cdot \v v^{(2)}
\right ).
\label{eq:O5-WZW-copies}
\end{equation}
We assume ``FM'' coupling between the N\'eel order parameters ($J_n < 0$)
and ``AFM'' coupling between the VBS order parameters ($J_v > 0$) %.The
\c{ so that the low energy configurations %would be
have}
$\v n^{(1)}=\v n^{(2)}\equiv \v n$ and $\v v^{(1)}=-\v v^{(2)}\equiv \v v$.
The action in terms of $\hat \spinfive=(\v n,\v v)$ will be similar to \Eq{eq:O5-WZW} but with a \emph{doubled} WZW term.

Note however $\v v=\frac{1}{2}(\v v^{(1)}-\v v^{(2)})$ cannot be directly measured in spin-1 systems,
because all physical observables must be symmetric with respect to
exchange of the two spin-1/2 moments.
Define two physical order parameters,
\begin{equation}
v'_1 = -\frac{v^{(1)}_x v^{(2)}_x - v^{(1)}_y v^{(2)}_y}{\sqrt{v_x^2+v_y^2}},\quad
v'_2 = -\frac{v^{(1)}_x v^{(2)}_y+v^{(2)}_x v^{(1)}_y}{\sqrt{v_x^2+v_y^2}}.
\label{vprime}
\end{equation}
$v'_1$ carries lattice momentum $(0,0)$, belongs to the $B_1$ representation of $C_{4v}$
(changes sign under 4-fold rotation,
but has no sign change under principal axis reflection),
and corresponds to the nematic order parameter $\Omega_4$ defined in the main text
(for example, the parent Hamiltonian ground state $|\text{X}\>$ in the main text 
corresponds to $v^{(1)}_x=-v^{(2)}_x\neq 0$ and $v^{(1)}_y=-v^{(2)}_y=0$ and thus $v'_1>0$). 
$v'_2$ has lattice momentum $(\pi,\pi)$, is the $B_2$ representation of $C_{4v}$
(changes sign under both 4-fold rotation and principal axis reflection 
around a lattice site),
and corresponds to 
certain superpositions of 
plaquette valence bond solid order
(for example, $v'_2<0$ may correspond to either $v^{(1)}_x=v^{(2)}_y>0$ 
with plaquette singlets centered at $(2x+1/2,2y+1/2)$,
or $v^{(1)}_y=v^{(2)}_x < 0$ with plaquette singlets centered at $(2x-1/2,2y-1/2)$ for integer $x,y$).
% {\it \B{This latter is not immediately clear to me}}
If we parametrize $\v v$ by $(v_x,v_y)=(v\cos\theta,v\sin\theta)$,
then $\v v'=(v'_1,v'_2)=(v\cos 2\theta,v\sin 2\theta)$.
Note that
\begin{equation*}
\begin{split}
&
v'_1\partial_\mu v'_2-v'_2\partial_\mu v'_1
=2v^2\partial_\mu\theta=2\,(v_x\partial_\mu v_y - v_y\partial_\mu v_x),
\\
&
\partial_\mu v'_1\,\partial_\nu v'_2
-\partial_\mu v'_2\,\partial_\nu v'_1
=2v\,(\partial_\mu v\, \partial_\nu \theta
-\partial_\mu \theta\, \partial_\nu v)
=2\,(\partial_\mu v_x\,\partial_\nu v_y
-\partial_\mu v_y\,\partial_\nu v_x).
\end{split}
\end{equation*}
Therefore the WZW term in action $S_1$ in terms of $\hat \spinfive'\propto
(\v n,\v v')=(n_x,n_y,n_z,v'_1,v'_2)$
has a \emph{halved} coefficient compared to that in terms of $(\v n,\v v)$.
The action then becomes
\begin{equation}
\action_{1}[\hat \spinfive^{(1)},\hat \spinfive^{(2)}]\sim \action_{1}[\hat \spinfive']=~
\action_{O(3)\times Z_2\times Z_2}[\hat \spinfive']
-2\pi \imth
\frac{3}{8\pi^2}
\int \dif u  \dif^2 x \dif\tau\,
\epsilon^{a b c d f}
\spinfive'_a
\partial_x \spinfive'_b
\partial_y \spinfive'_c
\partial_\tau \spinfive'_d
\partial_u \spinfive'_f.
\label{eq:O5-WZW-spin1}
\end{equation}
The action
$\action_{O(3)\times Z_2\times Z_2}[\hat \spinfive']$
is derived from $\action_{O(3)\times C_{4v}}[\hat{\phi}]$ while taking \Eq{vprime} into account.
It has $O(3)\times Z_2\times Z_2$ anisotropy induced by the additional anisotropy terms such as
$\Delta_{4}^{\vphantom{'}}\cdot \spinfive'^2_4$
and
$\Delta_{5}^{\vphantom{'}}\cdot \spinfive'^2_5$ where
$\Delta_{4,5}>0$.

Assume $\Delta_5 \gg \Delta_4 > 0$.
Consider the configuration
\begin{equation*}
\hat \spinfive'(u,x,y,\tau)
=\left (\hat \Omega(x,y,\tau)\sin(u),\cos(u) \right ),
\end{equation*}
where $\hat\Omega$ is the 4-component real vector field
in Eq.~(6) of the main text.
When $u=0$ this is a uniform space-time configuration,
and when $u=\pi/2$ this will be a low energy space-time configuration
with $\spinfive'_5=0$.
Integrate over $u$ from $0$ to $\pi/2$,
the WZW term becomes
\begin{equation*}
\begin{split}
&
-2\pi \imth
\frac{3}{8\pi^2}
\int \dif^2 x \dif\tau\,
\epsilon^{a b c d}
\Omega_a
\partial_x \Omega_b
\partial_y \Omega_c
\partial_\tau \Omega_d
\cdot
\int_0^{\pi/2} \dif u\,
(-\sin^5 u-\sin^3 u\cos^2 u)
\\
=~
&
-2\pi \imth
\frac{3}{8\pi^2}
\int \dif^2 x \dif\tau\,
\epsilon^{a b c d}
\Omega_a
\partial_x \Omega_b
\partial_y \Omega_c
\partial_\tau \Omega_d
\cdot
(-2/3)
\\
=~
&
\imth
\frac{\pi}{2\pi^2}
\int \dif^2 x \dif\tau\,
\epsilon^{a b c d}
\Omega_a
\partial_x \Omega_b
\partial_y \Omega_c
\partial_\tau \Omega_d
\end{split}
\end{equation*}
In the above the $(-\sin^5 u)$ and $(-\sin^3 u\cos^2 u)$ terms are
respectively from $abcdf=abcd5$ and $abcdf=5bcda$ terms in the WZW model
(terms with index ``$5$'' at other positions vanish).
This result is exactly the $\Theta$-term in
equation (6)
of the main text with $\Theta=\pi$.
Thus the final effective action is given by equation (6) of the main text. 
(Note that due to the anisotropy term $V(\Omega_4^2)$ the
stiffness constant for the first three components of $\hat{\Omega}$ will be different 
from that of $\Omega_4$ at low energies and long wavelengths.)

In the above discussion we have made the assumptions
that $J_n < 0$ and $J_v>0$ in \Eq{eq:O5-WZW-copies},
and $\Delta_5 \gg \Delta_4 > 0$ in \Eq{eq:O5-WZW-spin1}.
Here we briefly comment on several other possibilities.

(i). If $J_n <0$ and $J_v < 0$,
the low energy configurations in \Eq{eq:O5-WZW-copies}
will be $\v n^{(1)}=\v n^{(2)}\equiv \v n$ and $\v v^{(1)}=\v v^{(2)}\equiv \v v$,
this theory would describe the phase transition from
N\'eel AFM order ($\v n$) to columnar VBS order ($\v v$)
with a WZW term similar to that of \Eq{eq:O5-WZW} but with doubled coefficient.

(ii). If $J_n > 0$,
the low energy configurations in \Eq{eq:O5-WZW-copies}
would have $\v n^{(1)}=-\v n^{(2)}\equiv \v n$,
then $\v n$ is the director of the uniaxial spin-nematic order
(uniaxial spin-nematic states have $|\v n\cdot \v S=0\rangle$).
This theory would describe the phase transition
between ferro-spin-nematic order
and columnar VBS order (if $J_v < 0$) or a nematic quantum paramagnet (if $J_v > 0$).
However the WZW term would be absent,
so we expect this phase transition to be of first-order.

(iii). If $J_n<0$ and $J_v>0$ as we assumed in \Eq{eq:O5-WZW-copies},
but $\Delta_4\gg \Delta_5 > 0$ in \Eq{eq:O5-WZW-spin1}.
This theory would describe the phase transition
between N\'eel order and plaquette VBS order.

\section{The effects of explicit C$_4$ $\ra$ C$_2$ symmetry breaking field.}
\label{sec:RGflow}

To explicitly break the C$_4$ symmetry to C$_2$, we can introduce a $Z_2$
``Zeeman'' field
\begin{equation}
S\ra S+ h\int \dif^2 x \dif \tau~\Omega_4(x,y,\tau),
\label{zeeman}
\end{equation}
which breaks the degeneracy between the two types of monopole in Eq.(7) of the main text,
making the absolute value of the Feynman amplitude associated them different  -- hence they do not cancel.
Now tunneling events involving unit changes in the skyrmion number are allowed,
which causes mixing between the even and odd skyrmion sectors
which lifts the ground-state degeneracy.

\begin{center}
\begin{figure}
\includegraphics[scale=.65]{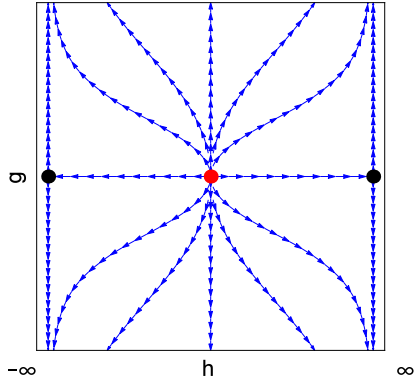}
\caption{
{\bf The renormalization group flow diagram of
\Eq{zeeman}.
The fixed points value of $g$ in this figure are all $O(1)$.
}
The proposed renormalization group flow diagram of
\Eq{zeeman}.
Along the $h=0$ axis the red critical point separates the small $g$ N\'eel ordered state
from the large $g$ nematic paramagnetic state.
Such transition if continuous will be an example of Landau-forbidden transition.
The blue critical point separates the anisotropic N\'eel state
(a state with N\'eel long range order but $\<\v S_{i}\cdot\v S_{i+\hat{x}}\>\ne \<\v S_{i}\cdot\v S_{i+\hat{y}}\>$)
and the anisotropic paramagnetic phase. Such phase transitions should belong to the $O(3)$ universality class.
}
\label{figS1}
\end{figure}
\end{center}

\begin{center}
\begin{figure}
\includegraphics[scale=.5]{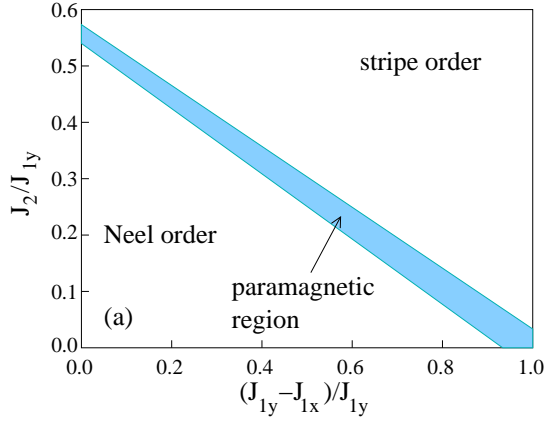}
\caption{
{\bf The phase diagram of the $J_1$-$J_2$ model, Eq.~1 of the main text.}
A schematic reproduction of the phase diagram reported in \Ref{JiangHC-PRB09}.
The line of $J_{1y}-J_{1x}=0$ shows the phase diagram for the $J_1$-$J_2$ model, Eq.~1 of the main text, with fourfold rotation symmetry.}
\label{figS2}
\end{figure}
\end{center}

Here we propose the renormalization group flow diagram
%[Fig.~1]
[\Fig{figS1}]
of \Eq{zeeman}.
Along the vertical axis at $h=0$ there is a continuous (Landau-forbidden), or weakly first order, phase transition
between the small $g$ N\'eel state and the large $g$ two-fold degenerate nematic PM phase.
This is supported by Fig.~4 of \Ref{JiangHC-PRB09}.
Along the axis where the absolute value of $h$ is large
the transition from the anisotropic N\'eel state (where there is N\'eel long-range order
but $\<\v S_{i}\cdot\v S_{i+\hat{x}}\>\ne \<\v S_{i}\cdot\v S_{i+\hat{y}}\>$) into the PM state
[this transition happens at the lower boundary of the diagonal strip in 
\Fig{figS2} discussed below]
is in the usual $O(3)$ universality class.
In the  large $g$ region of
%Fig.~1,
\Fig{figS1},
where the system remains PM, tuning $h$ from negative to positive encounters
a first order phase transition at $h=0$ while maintaining a nonzero $S=1$ gap.

Returning to the numerical results of \Ref{JiangHC-PRB09}, note that Jiang {\etal}
did study the effect of explicit rotation symmetry breaking on their results by
introducing anisotropy in the NN exchange constants in the $x$ and $y$ directions,
so that $J_{1y}>J_{1x}$.
They found for $0\le h\equiv (J_{1y}-J_{1x})/J_{1y}\le 1$ there is always
(for some range of $J_2/J_{1y}$) an intermediate PM phase
between the N\'eel and stripe ordered phases [see
% Fig.~2
\Fig{figS2}
for a schematic illustration].
Within this PM phase, they found no evidence of a phase transition, suggesting that it is all one phase.
Remarkably it is found that not only does this PM phase survive for $0.525 \lesssim J_2/J_{1}\lesssim0.555$ in the isotropic limit $h\to 0$,
but it also includes the case $J_2=0$ and $h=1$,
where the system consists of a decoupled array of spin-1 chains.
As it is independently known that the spin-1 AF chain is in the same phase as the spin-1 AKLT chain,
this observation nicely connects the results of the $J_1-J_2$ model to those obtained for $H_K$.

Finally, the other interesting scenario\cite{XuCK-Ludwig}, namely, the gapless state of Eq.~(6) in main text at $\Theta=\pi$
can be viewed as the critical state between two different SPTs (one at $\Theta=0$ and the other $\Theta=2\pi$).
Presumably such critical state can be obtained by proliferating the domain walls in the nematic order parameter in our nematic PM phase.
Further works are definitely deserved.

\section{Exact diagonalization results for the the $S=1$ $J_1$-$J_2$ model}
\label{sec:numerical}

\begin{center}
\begin{figure}
\includegraphics[scale=1]{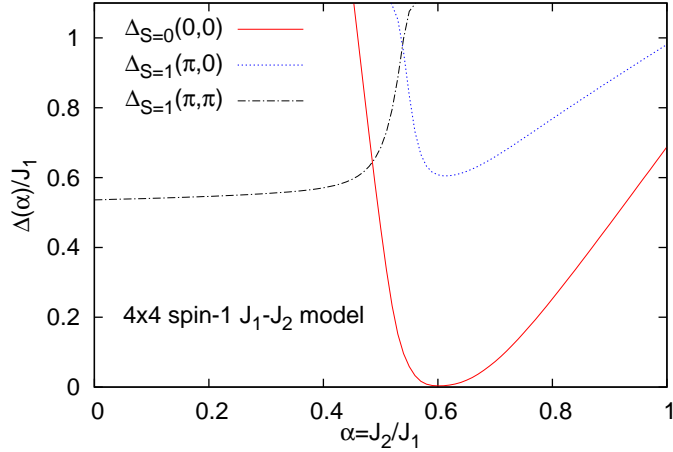}
\caption{
The global singlet($\Delta_{S=0}$) gap and the triplet($\Delta_{S=1}$) gaps at 
momenta $\vec{q}=(\pi,0)$ and $(\pi,\pi)$ for the spin-1 $J_1$-$J_2$ model
on $4\times 4$ lattice obtained by exact diagonalization.
The global triplet gap result is the minimum of the blue and black dashed curves. 
It exhibit a sharp kink consistent with the DMRG results of \Ref{JiangHC-PRB09}.
Due to the quantum tunneling between the two degenerate nematic PM states on finite lattices,
one expects an unique singlet ground state and a small gap for the singlet excitations.
The unique ground state for a given $\alpha=J_2/J_1$ is a spin singlet and has lattice momentum $(0,0)$.
The singlet gap is small when $J_2/J_1$ falls in the region where the nematic PM state exists
(it nearly vanishes around $\alpha=0.6$ where $\Delta_{S=0}(\alpha=0.6) = 0.0033 J_1$).
Moreover the lowest energy singlet excited state has lattice momentum $(0,0)$,
suggesting no translation symmetry breaking in the tentative nematic quantum PM state.
The triplet gap has a cusp around $\alpha= 0.54$.
The lowest energy $S=1$ states for $\alpha < 0.54$
have lattice momentum $(\pi,\pi)$ consistent with N\'eel order,
and the lowest energy $S=1$ states for $\alpha > 0.54$
have lattice momentum $(\pi,0)$ or $(0,\pi)$
consistent with stripe antiferromagnetic order.
}
\label{figS3}
\end{figure}
\end{center}

Exact diagonalization was performed on
$4\times 4$
%and $3\sqrt{2}\times 3\sqrt{2}$
square lattices with periodic boundary conditions
for the spin-1 $J_1$-$J_2$ antiferromagnetic Heisenberg model
\begin{equation}
H=J_1\sum_{\<ij\>}\v S_i\cdot\v S_j +J_2\sum_{\<\<jk\>\>}\v S_j\cdot\v S_k.
\label{eq:j1j2}
\end{equation}
The spin singlet and spin triplet gaps at momenta $\vec{q}=(\pi,0)$ and $(\pi,\pi)$ are presented for different values of $J_2/J_1$
in 
%Fig.~3. 
\Fig{figS3}
The global spin-1 gap is given by the minimum of the blue and dashed black curves. As a result it exhibits a kink consistent with that reported in
\Ref{JiangHC-PRB09}. What is noteworthy are (1) the spin-0 gap plunges in the range of $J_2/J_1$ where the nematic quantum PM state is expected. This 
presumably reflects the small splitting (due to quantum tunneling) between the two states that would be degenerate in the thermdynamic limit. 
(2) As the $(\pi,0)$ triplet gap vanishes as $J_2/J_1$ approaches the PM to stripe phase boundary, the $(\pi,\pi)$ triplet gap steadily increases. 
The reverse is true as $J_2/J_1$ approaches the PM to Neel phase boundary.
That within the PM regime, 
the $\Delta_{S=1}(\pi,\pi)$ is small compared to $\Delta_{S=1}(\pi,0)$ on the small $J_2/J_1$ 
and large on the large $J_2/J_1$ side 
suggests the existence of two closeby quantum phase transitions plays a key role in the physics.

\begin{center}
\begin{figure}
\includegraphics[scale=1]{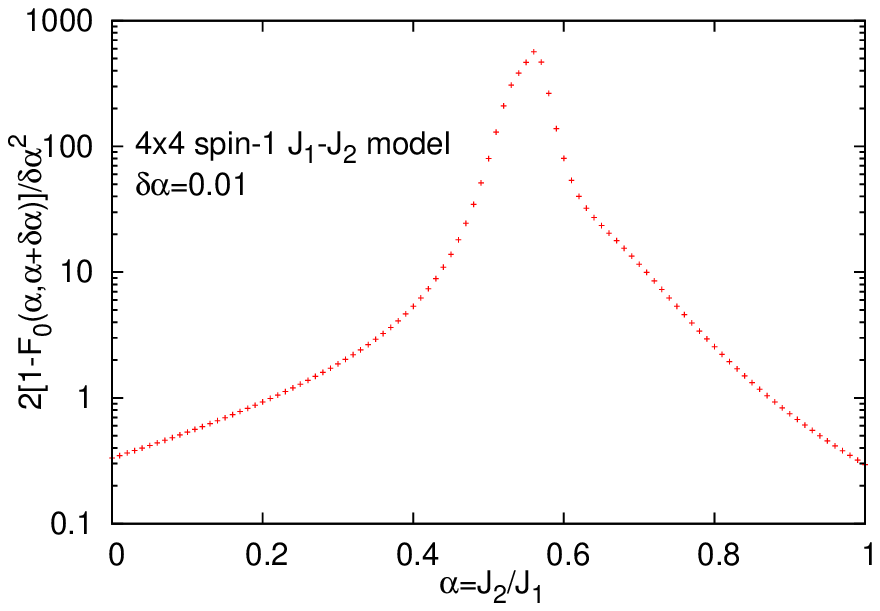}
\caption{
Ground state fidelity susceptibility for the spin-1 $J_1$-$J_2$ model on $4\times 4$ lattice.
The ground state fidelity $F_0(\alpha,\alpha+\delta\alpha)=
|\langle \psi_0(\alpha+\delta\alpha)|\psi_0(\alpha)\rangle|$
is the overlap between ground states ($\psi_0$) at parameter $\alpha$ and $\alpha+\delta\alpha$.
There is a sharp peak at around $\alpha=0.56$.
}
\label{figS4}
\end{figure}
\end{center}

The ground state fidelity susceptibility\cite{GuS-IJMPB10} is presented in 
%Fig.~4.
\Fig{figS4}.
This quantity displays a clear peak providing strong evidence of a quantum phase transition(s) within $0.5 < J_2/J_1 < 0.6$,
even on such a small lattice.
This result could be interpreted as indicating a single strongly first-order transition
(which is inconsistent with the result of \Ref{JiangHC-PRB09}).
Alternatively it can be taken as evidence of the existence of two (continuous) transitions (favored by the result of \Ref{JiangHC-PRB09}).

\section{Exact diagonalization results for a model interpolating between the parent Hamiltonian and $J_1$-$J_2$ model}
\label{sec:numerical-interpolating}

We consider a model which interpolates between the
parent Hamiltonian $H_K$ in Eq.~(2) of the main text 
and the $J_1$-$J_2$ Heisenberg model,
\begin{equation}
\begin{split}
H_{\lambda}
=
~&
\lambda\cdot \frac{15\,J_1}{4\,K}\cdot H_K+
(1-\lambda)\cdot (J_1\,\sum_{\langle ij\rangle} \v S_i\cdot \v S_j
+J_2\,\sum_{\langle\langle ij\rangle\rangle} \v S_i\cdot \v S_j)
\\
=~&
J_1\,\sum_{\langle ij\rangle} \v S_i\cdot \v S_j
+[J_2+\lambda\,(J_1/2-J_2)]\,\sum_{\langle\langle ij\rangle\rangle} \v S_i\cdot \v S_j
+\lambda\cdot (\text{higher order terms}),
\end{split}
\label{equ:interpolating}
\end{equation}
where the ``higher order terms'' contain those terms involving 4 or 6 spin operators.
This model becomes the parent Hamiltonian $\frac{15\,J_1}{4\,K}\cdot H_K$ at $\lambda=1$,
and the $J_1$-$J_2$ Heisenberg model at $\lambda=0$.
We study the behavior of this model for three
different $J_2/J_1$ values,
\begin{itemize}
\item
$J_2/J_1=0$: the $J_1$-$J_2$ Heisenberg model in this case
should exhibit N\'eel order.
The results for the interpolating model are shown in
%Fig.~5
\Fig{figS5}
and
%Fig.~6.
\Fig{figS6}.
In particular the ground state fidelity susceptibility shown in 
%Fig.~6
\Fig{figS6}
has a prominent peak at around $\lambda=0.9$,
suggesting a phase transition from nematic paramagnet at $\lambda=1$
to N\'eel order at $\lambda=0$.

\begin{center}
\begin{figure}
\includegraphics[scale=1]{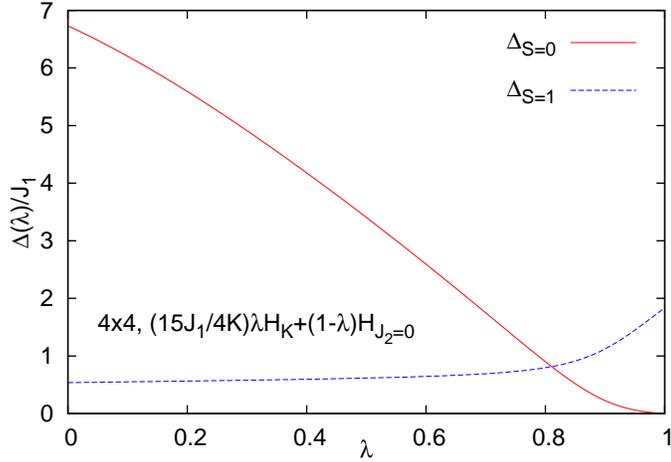}
\caption{
Singlet($\Delta_{S=0}$) and triplet($\Delta_{S=1}$) gaps for the spin-1 interpolating model \Eq{equ:interpolating} with $J_2=0$ on $4\times 4$ lattice.
}
\label{figS5}
\end{figure}
\end{center}

\begin{center}
\begin{figure}
\includegraphics[scale=1]{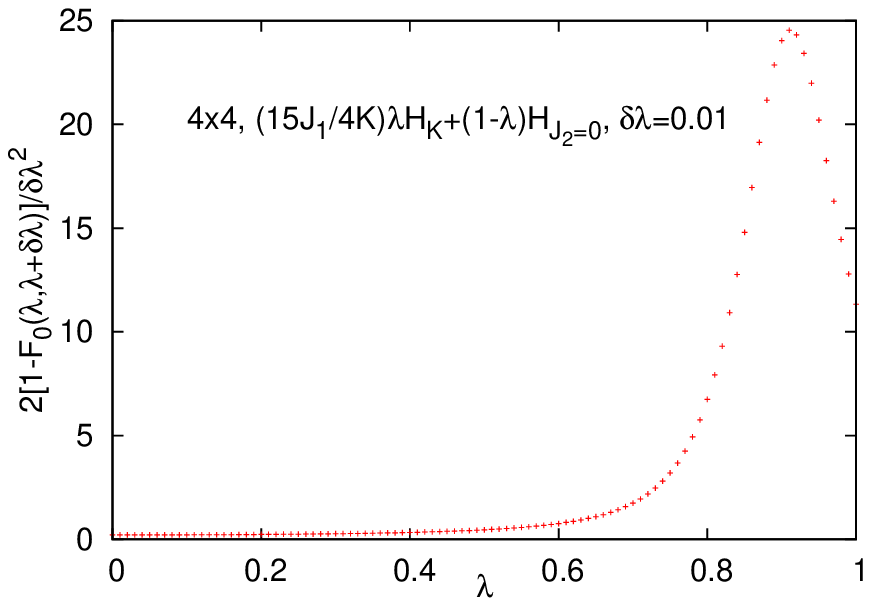}
\caption{
Ground state fidelity susceptibility for the spin-1 interpolating model \Eq{equ:interpolating} with $J_2=0$ on $4\times 4$ lattice.
There is a peak at around $\lambda=0.9$,
suggesting 
that this marks the transition point of a nematic paramagnetic phase for $\lambda > 0.9$ 
to a N\'eel ordered state for $\lambda < 0.9$.
}
\label{figS6}
\end{figure}
\end{center}

\item
$J_2/J_1=0.54$: according to the DMRG result in \Ref{JiangHC-PRB09},
the $J_1$-$J_2$ Heisenberg model in this case will be in PM phase.
The results for the interpolating model are shown in
%Fig.~7
\Fig{figS7}
and
%Fig.~8.
\Fig{figS8}.
The ground state fidelity susceptibility shown in 
%Fig.~8
\Fig{figS8}
has no peak,
suggesting that
the nonmagnetic phase at $\lambda=0$ ($J_2/J_1=0.54$) is
also a nematic paramagnet as the ground states of $\lambda=1$ parent Hamiltonian.

\begin{center}
\begin{figure}
\includegraphics[scale=1]{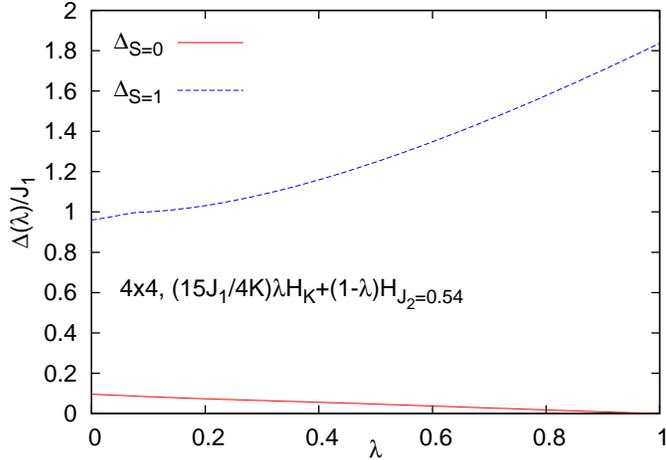}
\caption{
Singlet($\Delta_{S=0}$) and triplet($\Delta_{S=1}$) gaps for the spin-1 interpolating model \Eq{equ:interpolating} with $J_2/J_1=0.54$ on $4\times 4$ lattice.
The singlet gap remains small, suggesting that the nematic paramagnet phase
persists to $J_1$-$J_2$ model limit ($\lambda=0$).
}
\label{figS7}
\end{figure}
\end{center}

\begin{center}
\begin{figure}
\includegraphics[scale=1]{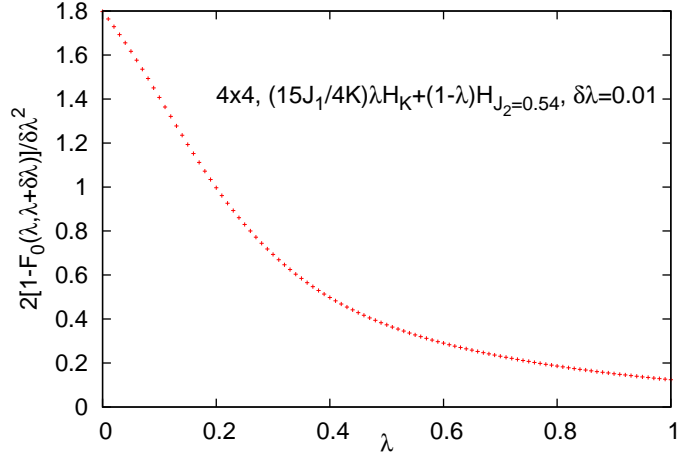}
\caption{
Ground state fidelity susceptibility for the spin-1 interpolating model \Eq{equ:interpolating} with $J_2/J_1=0.54$ on $4\times 4$ lattice.
There is no peak,
suggesting no phase transition from nematic paramagnet at $\lambda=1$ to the nonmagnetic phase at $\lambda=0$.
}
\label{figS8}
\end{figure}
\end{center}

\item
$J_2/J_1=1$: for this parameter choice the $J_1$-$J_2$ Heisenberg model should be deep inside
the stripe magnetic ordered phase.
The results for the interpolating model are shown in
%Fig.~9
\Fig{figS9}
and
%Fig.~10.
\Fig{figS10}.
The ground state fidelity susceptibility shown in 
%Fig.~10
\Fig{figS10}
has a prominent peak at around $\lambda=0.8$,
suggesting a phase transition from nematic paramagnet at $\lambda=1$
to stripe order at $\lambda=0$.

\begin{center}
\begin{figure}
\includegraphics[scale=1]{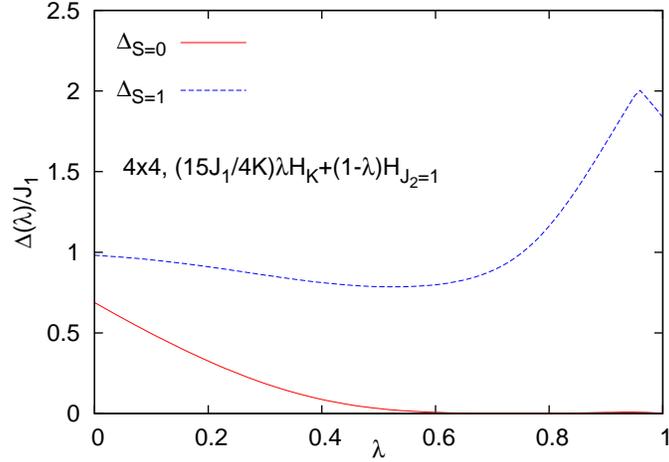}
\caption{
Singlet($\Delta_{S=0}$) and triplet($\Delta_{S=1}$) gaps for the spin-1 interpolating model \Eq{equ:interpolating} with $J_2/J_1=1$ on $4\times 4$ lattice.
The singlet gap remains small for $\lambda > 0.8$, suggesting that the nematic paramagnet phase
persists in this region.
}
\label{figS9}
\end{figure}
\end{center}

\begin{center}
\begin{figure}
\includegraphics[scale=1]{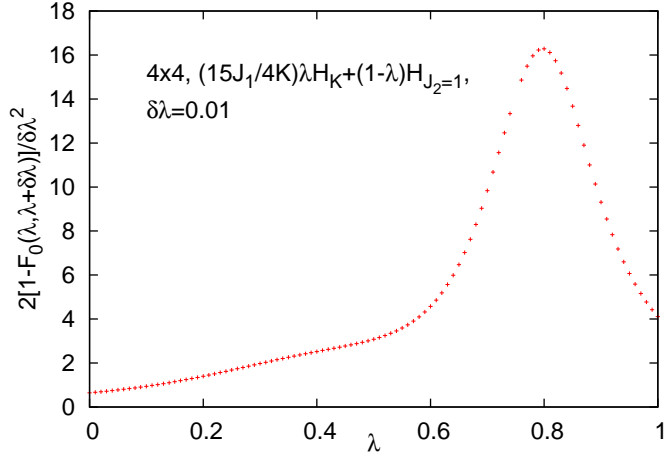}
\caption{
Ground state fidelity susceptibility for the spin-1 interpolating model \Eq{equ:interpolating} with $J_2/J_1=1$ on $4\times 4$ lattice.
There is a peak at around $\lambda=0.8$,
that this marks the transition point of a nematic paramagnetic phase for $\lambda > 0.8$ 
to a stripe ordered state for $\lambda < 0.8$.
}
\label{figS10}
\end{figure}
\end{center}

\end{itemize}

%% Put the bibliography here, most people will use BiBTeX in
%% which case the environment below should be replaced with
%% the \bibliography{} command.

% \bibliography{nematic}

% \begin{thebibliography}{1}
% \bibitem{dummy} Articles are restricted to 50 references, Letters
% to 30.
% \bibitem{dummyb} No compound references -- only one source per
% reference.
% \end{thebibliography}

%%
%% TABLES
%%
%% If there are any tables, put them here.
%%

\end{document}